\documentclass[aps,prl,10pt,twocolumn,showpacs,superscriptaddress,groupedaddress]{revtex4-2}

\usepackage{graphicx}
\usepackage{epstopdf}
\usepackage[normalem]{ulem}
\usepackage{amssymb}
\usepackage{dcolumn}
\usepackage{bm}
\usepackage{eqnarray}
\usepackage{ulem} 
\usepackage{amsmath}
\usepackage{float}
\usepackage{stackengine}
\usepackage{gensymb}
\usepackage{array}
\usepackage[]{subfigure}
\usepackage[dvipsnames]{xcolor}
\usepackage[geometry]{ifsym}
\usepackage{bm}
\newcommand{\RNum}[1]{\uppercase\expandafter{\romannumeral #1\relax}}

\usepackage{soul}
\usepackage{color}
\newcommand{\blue}[1]{\textcolor{black}{#1}} 
\newcommand{\red}[1]{\textcolor{black}{#1}}
\begin{document}

\preprint{APS/123-QED}

\title{Corona splashing triggered by a loose monolayer of particles}
\author{Ehsan Esmaili$^{1,3}$, Zih-Yin Chen$^2$, Anupam Pandey$^{1}$, Seungho Kim$^{1}$, Sungyon Lee$^{2}$ and Sunghwan Jung$^{1}$} \email{sunnyjsh@cornell.edu} \affiliation{1 Department of Biological and Environmental Engineering, Cornell University, Ithaca, NY 14853, USA \\
2 Department of Mechanical Engineering, 
University of Minnesota, Minneapolis, MN 55455, USA \\
3 Department of Mechanical Engineering, 
Purdue University, West Lafayette, IN 47907, USA}

\begin{abstract} 
In nature, high-speed raindrops often impact and spread on particulate surfaces (e.g., soil, plant leaves with spores or pollen). We study the dynamics of droplet impact on a loosely packed monolayer of particles by combining experimental and mathematical approaches. We find that the presence of mobile particles lowers the critical impact velocity at which the droplet exhibits corona splashing, as the particle area fraction is systematically increased. We rationalize this experimental observation by considering \blue{the jamming} of frictional particles at the spreading rim. Elucidating the splashing transition of the drop on a particulate bed can lead to a better understanding of soil loss and erosion from falling raindrops. 
\end{abstract}

\maketitle
\date{\today}

The dynamics of drop impact on loose solid particles (e.g., dust, soil, spores, and pollen) is important in understanding many engineering and natural processes, including  surface cleaning \cite{barthlott1997purity}, soil erosion by raindrops \cite{ahn2013effects,al1981new}, heat transfer \cite{liu2018droplet,kim2020raindrop}, and particle or waterborne disease transport due to rain impacts \cite{kim2019vortex,curriero2001association}.  
\begin{figure}
\centering
  \includegraphics[width=0.45\textwidth]{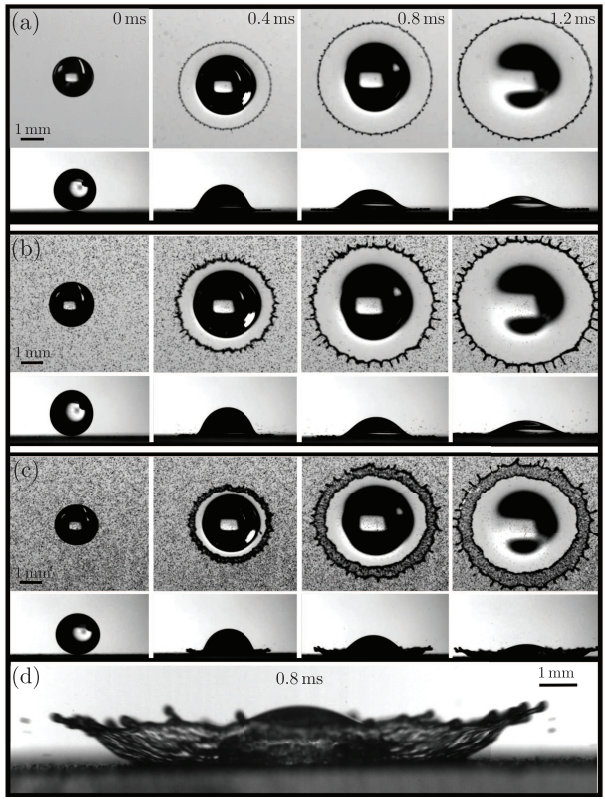}
    \caption{Bottom and side views of a drop \blue{(We $\approx$ 800)} impacting a glass surface with or without particles. (a) The drop spreads smoothly on the surface without particles. (b) On the glass surface with particles (packing fraction; $\phi\approx 0.13$), the drop still spreads smoothly with more pronounced rim instability. (c) With a higher packing fraction of glass particles ($\phi\approx 0.27$), the drop exhibits corona splashing. (d) Side view of corona splashing with high packing fraction ($\phi\approx 0.4$). }
    \vspace{-11 pt}
    \label{ExpTran}
\end{figure}
On a smooth surface without particles as in classical long-standing experiments, three regimes were observed \cite{rioboo2002time,pasandideh1996capillary,clanet2004maximal,josserand2016drop,chandra1991collision,eggers2010drop,roisman2009inertia,riboux2014experiments}: spreading, prompt splashing, and corona splashing. In the spreading regime, a thin liquid lamella spreads on the smooth surface without generating secondary droplets at the spreading front. In the prompt splashing regime, the spreading lamella becomes unstable and ejects small daughter droplets at the edge of the lamella \cite{yarin2006drop}. Then, corona splashing occurs at a higher impact speed, in which the spreading lamella detaches from the surface earlier and forms the crown-shaped lamella up in the air \cite{yarin1995impact,yarin2006drop,rioboo2003experimental,mundo1995droplet,burzynski2020splashing}. 

\begin{figure*}[t]
  \includegraphics[width=1\textwidth]{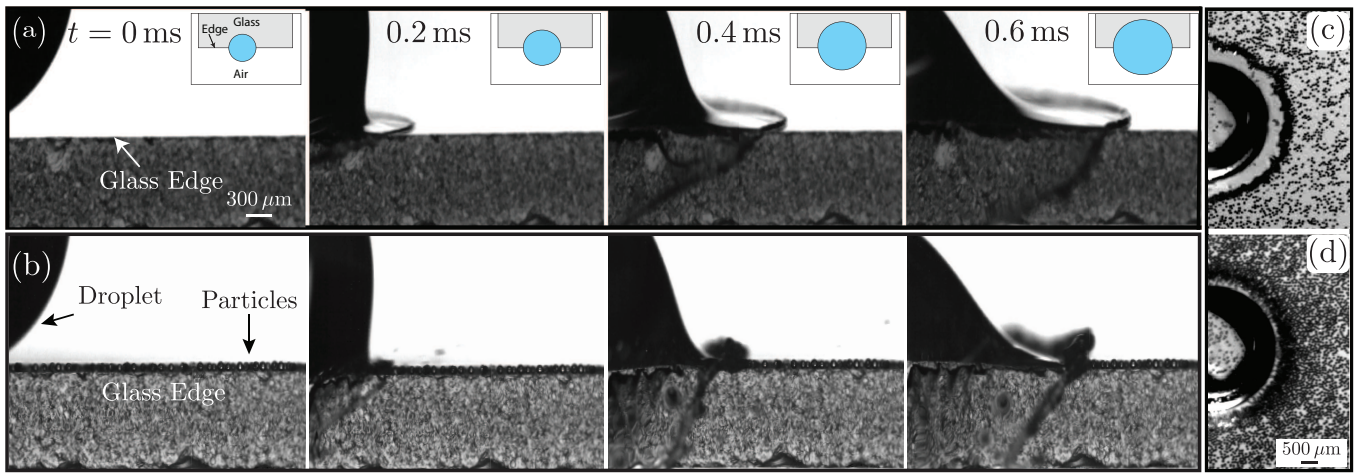}
   \caption{\blue {(a) Side view of a water drop (We = 612) spreading over the edge of the smooth glass surface with no particles. Insets show schematics of top view as the drop spreads. (b) A water drop spreads and splashes on the edge of the glass surface covered with glass particles ($r_p=50~\mu$m $\& $ $\phi\approx 0.33$ ). (c) A jamming front forms ahead of the rim for a water drop (We=450) impacting a particle layer of low packing fraction ($r_p=50~\mu$m $\& $ $\phi\approx 0.17$). (d) A jamming front forms on a particle layer of high packing fraction ($r_p=50~\mu$m $\& $ $\phi\approx 0.42$).}} 
    \label{Edgenew}
\end{figure*}

On a non-smooth surface with a particle bed, the splashing dynamics becomes more complicated and is affected by the properties of the particles \cite{delon2011impact,emady2013modeling,long2014experimental,zhao2017liquid}.
Previously, the inertia effect characterizing the shape and size of the crater \cite{katsuragi2010morphology}, the effect of the packing fraction \cite{marston2010experimental}, and the hydrophobicity \cite{supakar2016spreading} were investigated to elucidate the lamella interaction with granular media. These previous studies mostly used a thick layer of granular materials. However, the drop impact on a thin layer of dust and particles that may slide on the surface is not well understood \cite{xiong2019droplet,zheng2019drop}. 
At the outset, one would anticipate that the introduction of particles would resist the spreading motion, causing the lamella to slow down uniformly. However, our experiments found that the presence of particles above a certain volume fraction causes the lamella to lift off from the surface in lieu of slowing down, resulting in an unexpected corona splashing.

In this paper, we study the impact of a liquid drop on a solid substrate covered with a granular monolayer. For a fixed impact velocity, experiments reveal a discontinuous transition from spreading to corona splashing due to a small volume fraction of particles. Beyond this critical particle coverage, any drop impacting the surface splashes almost instantaneously before reaching its maximum spreading radius, \blue{which holds only for a certain range of particle packing fractions (see fig.~\ref{ExpTran}(a,b))}, in contrast to previous observations on granular impact~\cite{marston2010experimental}. By combining experiments and theoretical modeling, we elucidate the coupled dynamics of the liquid rim and the particles. We show that the particles during drop impact change the motion of the spreading rim and form a packed particle layer in front of the rim, which leads to corona splashing. A phase map of spreading versus corona splashing is presented and predicted by our model.

We conduct systematic experiments to observe how a liquid drop impacts and spreads over a monolayer of particles. We deposit a monolayer of micron sized glass particles on top of a glass slide by sieving through a steel 120 mesh. Particles of two different radii ($r_\mathrm{p}\!\approx\!$ 20 and 50\,$\mu$m) and specific gravity ($s_g\!\sim\!$ 2.5) are used in the experiments (see SI appendix Section A). The liquid drop is released from different heights by using a syringe pump (NE-1000 Programmable Single Syringe Pump), resulting in the impact velocity, $U$, in the range of 1-5 m/s and drop radius, $R_\mathrm{d}\approx$ \blue{1.8$\pm$0.1 mm}. In our study, Weber number (We=$2\rho_l U^2 R_\mathrm{d} / \sigma$) and Reynolds number (Re=$2\rho_l UR_\mathrm{d} / \mu$) range from 50 to 1250 and from 4000 to 20000, respectively. Two synchronized high speed cameras (Photron FASTCAM SA-Z and FASTCAM NOVA) are used to capture the bottom and side views of a spreading drop at a frame rate of 20,000-30,000 fps with a resolution of $1024 \times 672$ pixels. The bottom view images are analyzed using ImageJ software to calculate the ratio of the particle-occupied area to the total area. Then, the particle volume fraction, $\phi$, is determined by multiplying the areal ratio by $2/3$, \blue{ see SI appendix Section A. We note that the spatial variation of a particle packing fraction is quite low (1--4 $\%$). Thus we can assume a uniform packing fraction in experiments.}
 
Figure \ref{ExpTran} shows the side and bottom views of a water drop impacting the surface covered with glass particles of different volume fractions ($\phi=0, ~0.13, \mathrm{~and~} 0.27$) at approximately the same impact speed (We $\approx$ 800). Figure \ref{ExpTran}(a) displays the spreading dynamics on a clean glass surface without particles. Here, the lamella spreads smoothly over the surface until the rim reaches its maximum radius. At $\phi\approx 0.13$, the lamella spreads, and its rim becomes unstable and corrugated as shown in fig.~\ref{ExpTran}(b). 
By further increasing the particle packing fraction to $\phi\approx 0.27$, the corona splashing rim is observed as in fig.~\ref{ExpTran}(c,d). Thus, there clearly exists a critical particle packing fraction at which an impacting drop undergoes a transition from spreading to corona splashing. \blue{We define the corona splashing as a complete detachment of the rim from the surface (cf. fig.~\ref{ExpTran}(c)).} 

 \begin{figure*}[t]
  \includegraphics[width=.98\textwidth]{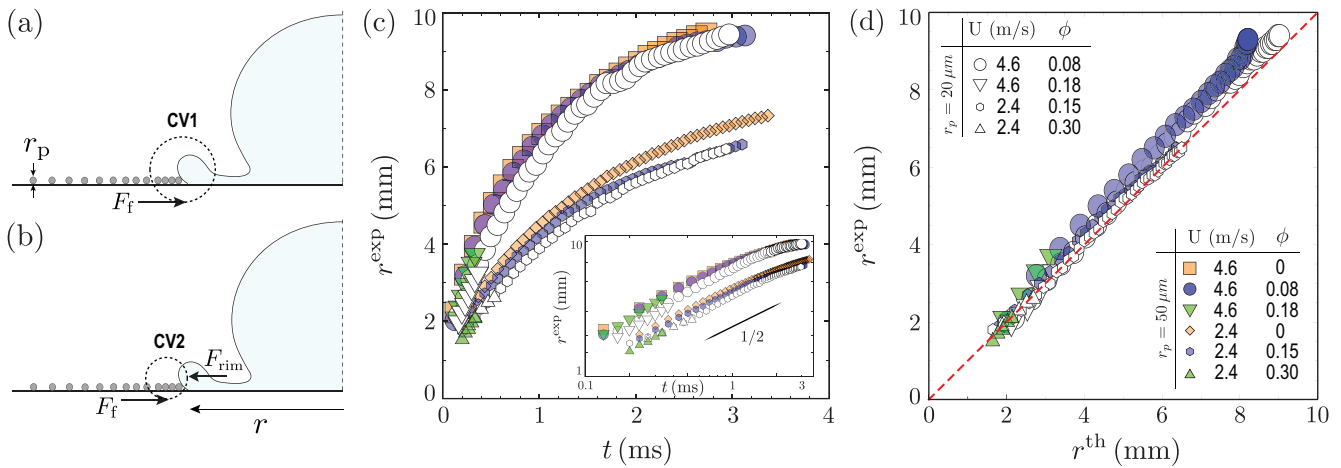}
   \caption{(a) The schematics of a spreading drop, showing the 2D control volume of the rim dynamics. \blue{CV1 includes the densely packed particles and the spreading rim whereas CV2 includes the densely packed particles only.} (b) The schematics of a spreading drop, including the control volume for \blue {jammed} particles only. (c) Spreading dynamics of a water drop for two different particle sizes. Inset shows that the rim spreading mostly follows the square root of time, $t^{1/2}$. (d) The comparison of the rim positions between experiments and our developed model, in which the slope of the dashed red line is 1.} 
    \label{spreading}
\end{figure*}

\begin{figure}
  \includegraphics[width=0.47\textwidth]{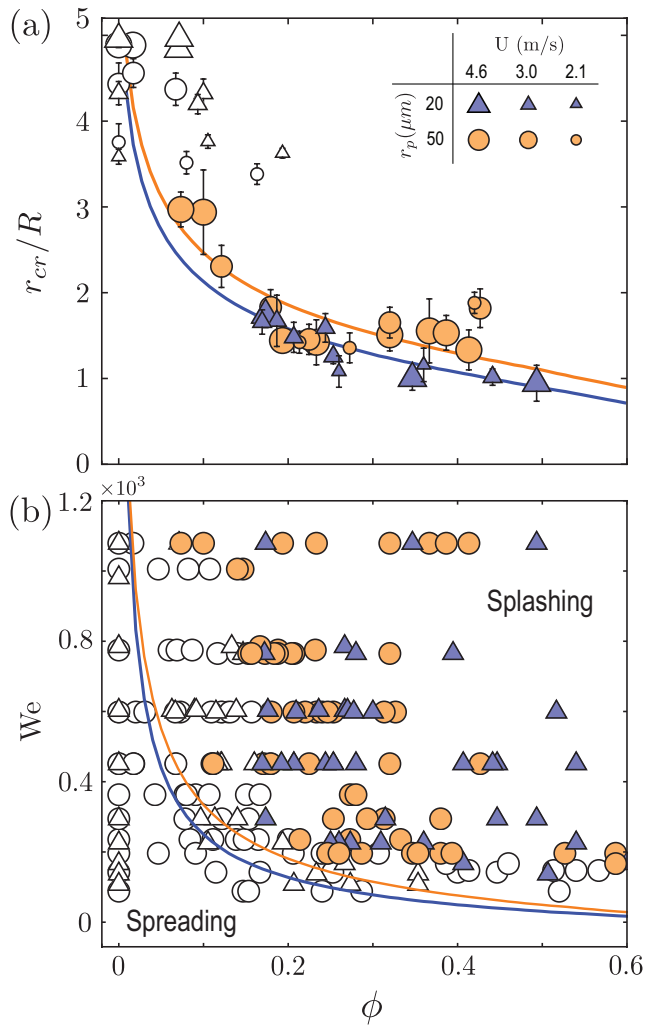}
    \caption{Transition from spreading to corona splashing. (a) Maximum radius of the rim attached to the surface, normalized by the drop radius ($R_d=1.8$ mm), for different packing fractions, particle sizes, and impact velocities. Open symbols represent spreading, whereas filled symbols are the corona-splashing cases. \blue{Solid lines represent theoretical predictions of the critical radius for 20 and 50 \,$\mu$m particles respectively}. (b) \blue{Phase map in terms of We number and the particle packing fraction, showing spreading-splashing behaviour of a water drop ($R_d=1.8 \pm 0.1$ mm)} on two different particle sizes. \blue{Solid lines show theoretical models extracted from Eq. \ref{phma2} for two particle sizes.}} 
    \label{Rcritic}
\end{figure}

To visualize the rim growth and particle-interface interactions in detail, we perform experiments of a drop impacting the edge of the particle-coated surface (see fig.~\ref{Edgenew}). \blue{Similar to experiments of a drop falling onto a semi-infinite surface \cite{chen2019drop},} 
we perform additional experiments to qualitatively probe the rim behavior. Figure \ref{Edgenew}(a) shows that without particles, the rim spreads along the surface, and its thickness does not grow rapidly. Then, at $\phi \approx 0.33$, the rim does not form immediately after the impact, as shown in fig.~\ref{Edgenew}(b). In contrast to the particle-free limit, the rim rapidly grows thicker and pushes the particles downstream. \blue{One may notice in fig.~\ref{Edgenew}(b) that the tip of the rim starts to levitate earlier than the detachment of the rim from the surface. This might lead to uncertainty in the measurement of the critical radius at splashing. We quantify and represent this uncertainty as error bars in fig.~\ref{Rcritic}(a) where the length of the error bars represents the difference in radii of the levitated and attached rims.} \blue{We speculate that particles can form the jammed front ahead of the rim (yet still being a monolayer), which are shown in fig.~\ref{Edgenew}(c,d). Even though a small portion of the particles are absorbed by the rim, the remaining particulate bed can redirect the spreading rim and lead to corona splashing.}

As evident in fig.~\ref{Edgenew}(b), a drop spreading on a particle-laden surface is a highly dynamic process. Upon droplet impact, the lamella and rim spontaneously form and spread radially outward, and the rim starts to collect and push the particles along the surface. Figures~\ref{spreading}(a) and \ref{spreading}(b) show the schematics of a spreading drop on a surface covered with particles. We plot the time-dependent position of the rim, $r(t)$, in fig.~\ref{spreading}(c), for varying impact velocities (i.e., $U\!=\!$ 2.4 or 4.6 m/s), particle packing fractions ($\phi\!=\!$ 0.08--0.30), and particle sizes ($r_\mathrm{p}\!=\!20\,\mu$m and $r_\mathrm{p} = 50\,\mu$m. The rim dynamics shows a weak dependency on the particle size and the particle packing fraction, but is primarily affected by the impact speed for both spreading and corona splashing cases. Moreover, the inset in fig.~\ref{spreading}(c) indicates the rim spreads approximately as the square root of time, $r \sim t^{1/2}$. This suggests that the rim dynamics is inertia-dominated, similar to previous studies~\cite{mongruel2009early,visser2015dynamics} in which $r \simeq (UR_\mathrm{d} \, t)^{1/2}$.

To elucidate the rim dynamics, we develop a 2D mathematical model in the manner of Gordillo et al.~\cite{gordillo2019} \blue{and incorporate the dynamic jamming front based on the work of Waitukaitis et al.~\cite{waitukaitis2013dynamic}} (SI appendix B and C). We compute the position $r(t)$ and the size $b(t)$ of the rim, by considering the mass and linear momentum of the spreading rim and the particles outside the rim. \blue{As evident in the top-down images of fig.~\ref{ExpTran}, most of the particles are displaced by the spreading droplet and become densely packed in front of the rim. To model the particle dynamics, we assume the particles are jammed at the advancing rim (i.e., $\phi_J=\pi/4$), while their overall spatial distribution, $\phi_{2\mathrm{D}}(x,t)$, is coupled to the speed of the rim and is computed with $r(t)$ and $b(t)$ \cite{waitukaitis2013dynamic}. \red{Note that $\phi_{2D}(x,0)=\phi$ is a constant representing the uniform particle distribution prior to the droplet impact.} We estimate the total 2D-mass of the jammed particles as $m_\mathrm{acc}(t) = N\rho_\mathrm{p}\pi r_\mathrm{p}^2$, where 
\red{$N\approx 2\phi x'(t)/(\pi r_\mathrm{p})$} is the number of densely packed particles; $x'(t)$ is the location at which $\phi_{2\mathrm{D}}<0.99\phi_J$ (see SI appendix Section B).}

We empirically model the frictional force on the particles from the substrate (fig.~\ref{spreading}(a,b)) as $F_f = \mu_f m_{\mathrm {acc}}(t)g$, where $g$ is the gravitational acceleration, and the dimensionless friction coefficient, $\mu_f$, depends on the dimensionless shear rate $\dot\gamma/\sqrt{g/r_{\mathrm{p}}}$ based on  \cite{daCruz2002,forterre2008} where $\dot\gamma\sim U/r_{\mathrm{p}}$. The choice of $\mu_f$ directly affects the magnitude of the resistant force, which also influences the threshold of droplet splashing. For instance, an increase in $\mu_f$ will tend to stunt the growth of $r(t)$, leading to a smaller maximum spreading radius. \blue{In addition, $\mu_f$ includes a fitting parameter, $\alpha$, whose value we adjust to compare the model with the experimental results. The sensitivity analysis of $\alpha$ is included in SI appendix Section B.}

Overall, figure \ref{spreading}(d) shows good agreement between the experimentally measured position of the rim and our model above, especially at early times (i.e., smaller $r^\mathrm{exp}$) when the fluid inertia dominates over the granular friction. Deviations between theory and experiments become more notable at later times (i.e., larger $r^\mathrm{exp}$), which may be attributed to the empirical nature of the frictional force, as well as the neglect of the viscous effects.

At low packing fractions, the rim stays on the surface until it reaches its maximum spreading radius (open symbols; spreading regime) in fig.~\ref{Rcritic}(a). It is worth mentioning that $r_{cr}$ in fig.~\ref{Rcritic}(a) is defined as the maximum radius over which the rim stays on the solid surface. For \blue{non-splashing cases}, $r_{cr}$ is the same as the maximum spreading radius. As $\phi$ increases, the rim becomes unstable and exhibits corona splashing (filled symbols). The radius at the moment of the ejection of the spreading drop, $r_{cr}$ is less than $r_{cr}$ without particles. As previously noted, most of the particles in contact with a spreading drop are radially pushed out and \blue{jammed} in front of the rim (fig.~S3 in SI appendix Section D). \blue{The jammed particles resist the inertia of the spreading rim due to particle-particle and particle-surface friction.} Consequently, this packed particle bed redirects the rim slightly upward and induces the corona splashing, analogous to droplet splashing due to a small fixed obstacle on the surface \cite{josserand2005droplet}.

To rationalize this critical rim radius at which splashing initiates, we consider the linear momentum of the \blue{jammed} particles:
\begin{align}
    \frac{d}{dt}\left[m_{\mathrm {acc}}(t)u_{\mathrm p}\right] = F_{\mathrm rim}-F_f(t),
\end{align}
where $u_{\mathrm p}$ is the velocity of the packed particles, while $F_{\mathrm rim}$ corresponds to the force of the spreading rim on the \blue{jammed} particles. Distinct from the model discussed in the appendix, the control volume here includes the \blue{jammed} particles only, as shown in fig.~\ref{spreading}(b). Hence, $F_{\mathrm rim}$ is the force from the spreading rim that pushes the particles forward, while $F_f$ is the frictional force exerted on the particles by the substrate. Note that $F_{\mathrm rim}$ does not show up explicitly in the previous model that includes both the particles and the rim in its control volume. Prior to splashing, the rim and the particles are assumed to move together, so that $u_{\mathrm p} = \dot r(t)$. As time progresses, the granular frictional force $F_f$ grows with $m_{\mathrm acc}(t)$, exceeding $F_{\mathrm rim}$. When $F_f>F_{\mathrm rim}$, the \blue{jammed} particles no longer move with the rim and act as a rigid ``obstacle'' that may induce splashing, in a similar manner of \cite{josserand2005droplet}. Additionally, we observe that a drop corona-splashes at a larger angle of ejection with large particles compared to that with smaller particles (see SI appendix Section E). Hence, we set the critical radius of splashing, $r_{cr}$, for $F_f\!=\!F_{\mathrm rim}$, or $d\left[m_{\mathrm acc}(t)u_{\mathrm p}\right]/dt\!=\!0$. For the current simulations, we assume that all the displaced particles \blue{are packed} at the front of the rim, instead of being incorporated into it. 


As shown in fig.~\ref{Rcritic}(a), the resultant critical radius of splashing, $r_{\mathrm cr}$, is plotted as two solid lines for $r_{\mathrm p}$ = $20\,\mu$m (orange) and $50\,\mu$m (blue-violet), respectively. Consistent with the experimental results, $r_{\mathrm cr}$ only shows weak dependence on the impact speed, as each solid line is an overlay of three distinct values of $U$ ranging from 2.1 to 4.6 m/s. Our results also show that the particle size does not alter the maximum spreading radius significantly. The impact velocity strongly affects only the maximum radius for spreading drops, not the maximum radius for corona-splashing drops.

Overall, the match between theory and experiment is quite reasonable for $\phi >0.1$, where the droplet is observed to undergo splashing at $r_{cr}$ in the experiments. However, at low packing fractions (i.e., $\phi <0.1$), the rim stays on the surface until it reaches its maximum spreading radius (open symbols). Its maximum radius of the spreading drop can be approximated as \cite{gordillo2019theory}
\begin{equation}
r_\mathrm{max} = k R_\mathrm{d} \mathrm{We}^{1/2} \mathrm{~(for~smooth~surfaces~only)}, \label{spreading2}
\end{equation}
where $k\!=\!0.19$ (see SI appendix section F). Notably, our current theoretical model of $r_{cr}$ at which $d/dt\left[m_{\mathrm acc}(t)u_{\mathrm p}\right]=0$ is consistently lower than $r_\mathrm{max}$ for $\phi <0.1$. This deviation between theory and experiment is expected, as our simulations tend to overpredict the amount of \blue{jammed} particles and, hence, the frictional force opposing the rim growth.

Finally, the transition from spreading to corona splashing can be predicted by determining the crossing point between two predicted radii: the maximum spreading radius (Eq.~(\ref{spreading2})) and the the critical radius, $r_{\mathrm cr}$. The corresponding impact velocity becomes
\begin{equation}
U_\mathrm{transition} = \frac{r_{cr}(\phi,r_{\mathrm p})}{k R\sqrt{2\rho R/ \sigma}}. \label{phma2} 
\end{equation} 
Figure \ref{Rcritic}(b) shows the phase map of spreading (open symbols) vs. corona splashing (filled symbols) with two different particle sizes. The two solid lines correspond to the results of Eq.~(\ref{phma2}) for $r_{\mathrm p}$ = $20\,\mu$m (orange) and $50\,\mu$m (blue-violet), respectively, and are shown to effectively separate the experimental spreading and corona splashing cases. \blue{It should also be noted that there is a possibility of micro splashing while a high-speed drop impacts a smooth surface without particles \cite{thoroddsen2012micro}. However, we herein focus on macro-splashing where the whole liquid rim detaches from the surface and form a corona-shaped splash.} It is worth mentioning that we observe a similar phase map for a drop impacting on a hydrophobic surface (see SI appendix Section G), showing \blue{
that the corona splashing phenomenon is independent of surface wettability}. 

In this present study, we found that a small amount of particles on the surface can trigger the corona splashing via interactions between the rim and particles. As the drop impacts the surface covered with particles, the rim spreads as the square root of time, similar to high inertia spreading cases on a smooth dry surface. However, distinct from the particle-free case, the spreading rim is resisted by the granular particles that are packed in front of it, forming a densely packed particle bed. Consequently, the rim detaches from the surface, resulting in corona splashing, which is explained by our theoretical model. The results of our fundamental study suggest a strongly coupled dynamics between the droplet and particles, which has implications for various natural processes, such as waterborne disease transport and soil erosion.

See the supplementary material for details of experimental measurements and the model of spreading dynamics.\\

This work was supported by National Science Foundation (NSF) Grant CBET-1919753. 
This X-ray imagining used resources of the Advanced Photon Source led by Dr. Kamel Fezza and Dr. Tao Sun, a US Department of Energy (DOE) Office of Science User Facility operated for the DOE Office of Science by Argonne National Laboratory under Contract DE-AC02-06CH11357.\\ 

\textbf{DATA AVAILABILITY}\\
The data and findings of this study are available at http://doi.org/10.17605/OSF.IO/3CWT2

\bibliographystyle{apsrev4-1}
\bibliography{Refe.bib}
\end{document}